# Evaluation of the citation matching algorithms of CWTS and iFQ in comparison to Web of Science

Marlies Olensky[1, 2], Marion Schmidt[2], Nees Jan van Eck[3]

[1]*marlies.olensky@gmail.com*          [2]*schmidt@forschungsinfo.de*          [3]*ecknjpvan@cwts.leidenuniv.nl*
[1] Humboldt-Universität zu Berlin, Berlin School of Library and Information Science, Unter den Linden 6, 10099 Berlin (Germany), Tel. +886 987 967 046
[2] Institute for Research Information and Quality Assurance (iFQ), Schützenstr. 6a, 10117 Berlin (Germany), Tel. +49 30 206 41 77-32
[3] Centre for Science and Technology Studies (CWTS), Leiden University, P.O. Box 905, 2300 AX Leiden (The Netherlands), Tel. +31 71 527 3909

**Keywords**
Bibliometric research groups, CWTS, iFQ, citation matching, evaluation, data accuracy, missed citations, Web of Science

**Abstract**
The results of bibliometric studies provided by bibliometric research groups, e.g. the Centre for Science and Technology Studies (CWTS) and the Institute for Research Information and Quality Assurance (iFQ), are often used in the process of research assessment. Their databases use Web of Science (WoS) citation data, which they match according to their own matching algorithms – in the case of CWTS for standard usage in their studies and in the case of iFQ on an experimental basis. Since the problem of non-matched citations in WoS persists because of inaccuracies in the references or inaccuracies introduced in the data extraction process, it is important to ascertain how well these inaccuracies are rectified in these citation matching algorithms. This paper evaluates the algorithms of CWTS and iFQ in comparison to WoS in a quantitative and a qualitative analysis. The analysis builds upon the methodology and the manually verified corpus of a previous study. The algorithm of CWTS performs best, closely followed by that of iFQ. The WoS algorithm still performs quite well ($F_1$ score: 96.41%), but shows deficits in matching references containing inaccuracies. An additional problem is posed by incorrectly provided cited reference information in source articles by WoS.

## Introduction

Bibliometric indicators analyze different aspects of research and are often consulted to support decision-making when research funds are assigned. Efforts are invested into making the results of bibliometric calculations, i.e. bibliometric indicators, comparable and new ways to measure the impact and productivity of research are constantly being developed. The underlying data sources used in bibliometric studies are usually one or more citation indexes, such as Web of Science (WoS), Scopus and also, increasingly, Google Scholar. The data quality of the citation indexes has been discussed in comparative studies addressing coverage and overlap (e.g. Archambault, Campbell, Gingras, & Larivière, 2009; Meho & Yang, 2007). Additionally, researchers have developed a model that aims to estimate the missed citation rate of a specific corpus of publications by comparing two citation indexes with each other (Franceschini, Maisano & Mastrogiacomo, 2013a, b). However, what do we actually know about the quality of the underlying data of citation indexes – namely, the cited references and the ability of the citation matching algorithms to compensate for inaccurate data in the references?

Producers and users of a bibliometric study should know certain crucial aspects about the data used, for example, how the publication data of the original articles was collected or how the cited references were matched to their target articles (Moed, 2002). Hence, the reliability of a bibliometric study depends on the accuracy of citation counts. Citation counts, in turn, depend on the data accuracy of the bibliographic references as well as the citation matching process which ideally can compensate for any inaccurate data in the references and ensure a correct match.

Several studies have investigated the data quality of bibliographic references and the error rate in terms of non-matched, i.e. missed, citations in WoS and found error rates of 6-12%, depending on different aspects of the data samples (Larsen, Hytteballe Ibanez & Bolling, 2007; Tunger, Haustein, Ruppert, Luca & Unterhalt, 2010; Moed & Vriens, 1989; Hildebrandt & Larsen, 2008; Moed, 2005; Olensky, 2015). Even though sophisticated algorithms for matching cited references to their respective target articles have been developed by bibliometric research groups, inaccuracies in bibliometric data sources, like WoS and Scopus, still occur (Neuhaus & Daniel, 2008), leading to missed or incorrect matches. Hence, data quality problems in bibliometric data sources are far from being solved (Franceschini, Maisano & Mastrogiacomo, 2013b) and can ultimately be traced back to the bibliographic data in references, the data extraction process as well as citation matching algorithms (Olensky, 2015). To date, little information has been published about the citation matching algorithm of WoS (Larsen et al., 2007; Hildebrandt & Larsen, 2008) and even less about the algorithms developed by bibliometric research groups. Only one bibliometric research group, the iFQ in Berlin, has published parts of the development process of such a matching algorithm (Schmidt, 2012), which is not yet in production. Others have refrained from publishing their solutions, mainly to retain strategic advantages.

In accordance with good scientific practice, CWTS and iFQ decided to evaluate their citation algorithms, in order to demonstrate their effectiveness compared to WoS. In a previous study (Olensky, 2015), inaccuracies in manually verified bibliographic references missed by WoS were investigated and compared with different citation matching algorithms (those of WoS, Scopus, Google Scholar, CWTS, iFQ and Science-Metrix) in order to establish which of them was able to rectify inaccurate data best. The present research, however, expands the previous study by comparing the entire corpus of matches found by WoS, CWTS and iFQ in order to determine the recall and precision of the algorithms. Additionally, a comparison of the inaccuracies in citations missed by the three data sources can contribute towards further pinpointing rules that can be tested within the citation matching algorithms of the bibliometric research groups.

The paper is organized as follows: first, related work on inaccuracies in bibliographic data as well as the citation matching algorithms by the three data sources investigated is elaborated. The ensuing section describes the data sample and the applied methodology. Next the results of the quantitative and the qualitative analysis are presented. The penultimate section discusses the findings, followed by the conclusion.

## Related work

### Inaccuracies in bibliographic data

The most commonly reported inaccuracies in the references of individual papers are variations, inconsistencies and errors in author name, journal title, publication year, volume and starting page number (Moed & Vriens, 1989; Galvez & de Moya-Anegón; 2006; Galvez & de Moya-Anegón, 2007; Neuhaus & Daniel, 2008; Adriaanse & Rensleigh, 2013; Chang, McAleer & Oxley, 2011). Additionally, studies have noted an increased likelihood of errors in publications written by consortia, i.e. large groups of authors (Moed, 2002; van Raan, 2005), journals with dual volume-numbering systems or combined volumes, and journals applying different article-numbering systems (Moed, 2002; van Raan, 2005; Tunger et al., 2010). Another problem is posed by different language backgrounds, which can lead to a misunderstanding of author names (Sweetland, 1989; van Raan, 2005; Moed, 2005).

The most inaccurate bibliographic fields have also been investigated in various studies and, interestingly, all bibliographic fields are represented: article title, author name and publication year (Meho & Rogers, 2008); volume and page number (Jacsó, 2004); page number, author names and year (Hildebrandt & Larsen, 2008); volume number, followed by a double error in volume number and starting page number, and in only very few instances, a wrong starting page number (Liang, Zhong & Rousseau, 2014); article title followed by ending page number and author-related fields (Olensky, 2015). Consequently, it is not possible to identify "the" most inaccurate bibliographic field; it depends on the methodology applied, i.e. the definition of what an inaccuracy is and the fields investigated.

Error rates, understood as references with inaccuracies which resulted in missed matches in WoS, are reported to be between 5.6% and 12% (5.6%: Olensky, 2015; 6.2%: Larsen et al., 2007; 7%: Tunger et al., 2010; 9.4%: Moed & Vriens, 1989; 12%: Hildebrandt & Larsen, 2008). Differently sampled data corpora cause the rather high variation in error rates. Moed (2005) investigated 22 million citing references by employing different matching rules in order to match the references to their 18 million target articles, which is the most comprehensive study on the accuracy of citing references in WoS to date. He found 7.7% discrepant references, which resulted in a missed match in WoS. However, the definitions of an error, discrepancy and inaccuracy differ in all these studies (Olensky, 2012). The error rates are, therefore, not strictly comparable, but still permit an estimate that the average missed citation rate in WoS may range between 5% and 12%. In order to classify inaccuracies occurring in bibliographic data in a standardized way and make the error rates comparable, Olensky (2015) developed a taxonomy of bibliographic inaccuracies. The taxonomy (cf. Figure 1) consists of three levels, where the lowest level is composed of inaccuracy codes (IACs) that decode certain types of inaccuracies (e.g. K describes an inaccuracy code for a space character discrepancy [1]). The inaccuracies are summarized according to common characteristics in the middle level of the taxonomy, such as disarranged data values or spelling variations. The upper level of the taxonomy describes the sophistication of a rule required to transfer discrepant values into the correct values.

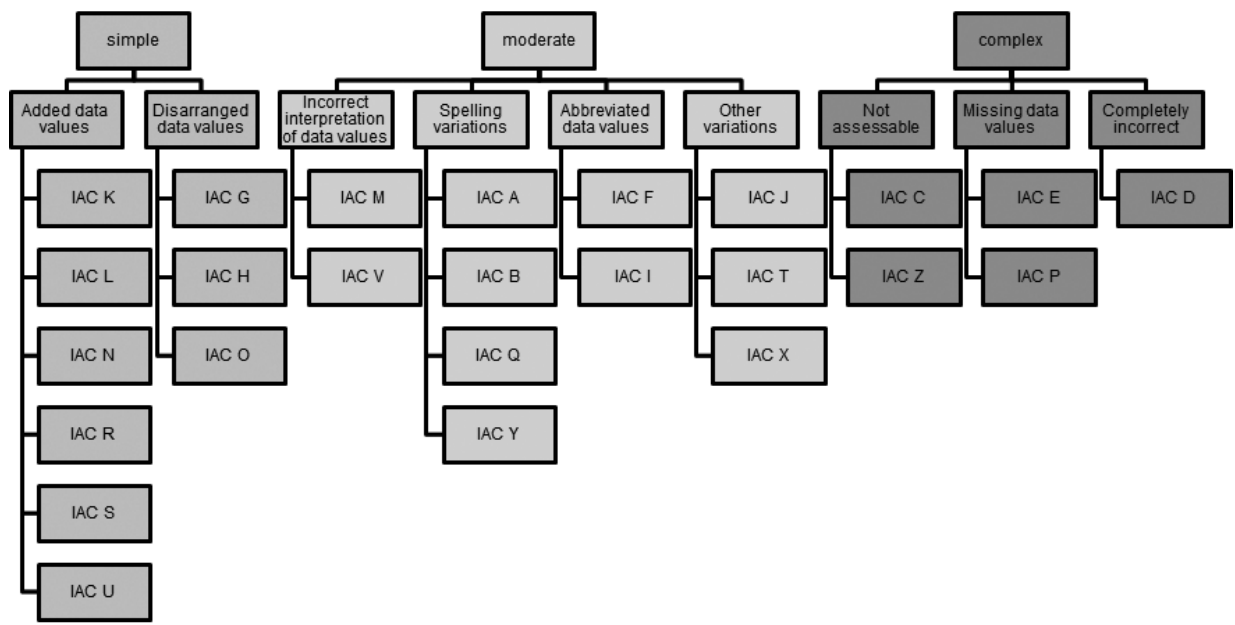

**Figure 1. Taxonomy of bibliographic inaccuracies (Olensky, 2015)**

Inaccuracies in bibliographic data can be caused either by the publishing author (e.g. provides inconsistent name variations including or excluding the second initial) or by the citing author (e.g. misspells one of the author names) (Olensky, 2015). Inaccuracies in references are sometimes corrected in the copy editing process by journal publishers (Meyer, 2008), any remaining inaccuracies in the data will be published. The data delivery, ingestion or extraction process in a citation index can additionally introduce inaccuracies (e.g. interprets the cited page number as the starting page number), but can also correct inaccurate data[2]. For a long time the metadata records in citation indexes were obtained from scanned documents (Moed, 2005) where the cited reference information was extracted using OCR software. The scanning process as such is error-prone, and different citation styles may also cause inaccuracies in this process (Meyer, 2008). In any case, the inaccuracies in the process may lead to a non-link between target and source article. With the transition to XML-formatted documents, publishers can provide metadata records directly to the citation index (Meyer, 2008; Moed, 2005), which can be expected to result in more accurate data.

*The citation matching process*

Citation matching is the process that matches the cited reference information from a source article to its cited, i.e. target, article. A citation that is not matched to a target article (even though the target article is available in the database) is usually called one of the following: missed citation, omitted citation, false negative match or stray reference. In this paper, such citations are referred to as missed matches. Cited references which are linked to target articles, where the original source article does not contain a reference to the respective target article, are referred to as incorrect matches, i.e. false positive matches. The accuracy of citation links is influenced by the process of data extraction from the publications as well as the accuracy of the cited reference information to be matched. Additionally, the citation matching process needs to be sufficiently sophisticated to find an optimal balance between missed and incorrect matches. Since no studies that evaluate different citation matching algorithms have been published to date, the citation matching process has been an invisible characteristic of bibliometric studies. Usually, a user of a citation analysis does not receive any information about the citation matching process applied, but simply has to rely on its accuracy.

The bibliographic data employed in citation matching was determined about 50 years ago by the first available source of bibliographic and citation data: the Science Citation Index, which evolved to the WoS. The Institute for Scientific Information (ISI) decided to extract the following information from the bibliography of an article to use it to link cited references to their respective target articles: first author, abbreviated source title (= publication name), year, volume and starting page number (Moed, 2005). Due to high data storage costs at the time,

ISI chose to extract only the first author from a citing reference and, therefore, was able to provide greater coverage of source titles (Garfield, 1990). Even though mass storage has become cheaper in the past few decades, Thomson Reuters, today's owner and operator of the WoS, did not change its policy for extracting cited references until 2012. Nevertheless, the cited reference information available for download from the WoS web interface has not been influenced by this change of extraction or storage policy.

*The citation matching algorithm of WoS*

Little is known about the citation matching algorithm used in WoS. As mentioned in the introduction, Thomson Reuters also refrains from revealing any details about the actual algorithm applied. Larsen et al. (2007) concluded from their investigation of missed matches in WoS that the algorithm must be quite conservative and not allow for any variations. In 2012, Thomson Reuters started to make more complete cited reference information (including more or the full list of authors, the complete publication name and the article title) available for not all, but some citations that are not automatically linked in WoS. However, downloading the bibliographic record of the source article still only gives the user the abbreviated format of cited references (first author, publication year, abbreviated publication name, volume number, starting page number and if available the DOI). The tagged data format which is available for subscribers of the raw data, such as iFQ and CWTS, provides the following bibliographic fields: first author, publication year, abbreviated publication name, volume number, starting page number, (if available) the DOI and sometimes the issue number. The XML format, which is provided by Thomson Reuters since 2013, seems mostly, but not fully, concordant with the full cited reference information in the online interface. Figure 2 shows two examples extracted from an XML dump from WoS provided to iFQ, which makes it clear that non-matched cited references of current publications in WoS-XML are not always delivered with all author names (because both publications contain multiple authors in the original publication) and do not always contain the article title either (in the first example concordant with the online interface). In contrast, Scopus provides the full cited reference information exactly as they extract it from the source articles as a download via their web interface.

```
<reference>³
<citedAuthor>REICHEL MP</citedAuthor>
<year>2013</year>
<citedWork>INT J PARAS IN PRESS</citedWork>
</reference>

<reference>⁴
<citedAuthor>Melbye, M</citedAuthor>
<year>2013</year>
<page>13</page>
<volume>33</volume>
<citedTitle>A first look: determinants of dental
care for children in foster care</citedTitle>
<citedWork>Spec Care Dentist</citedWork>
</reference>
```

**Figure 2. Examples of two cited references from an XML dump from WoS provided to iFQ as part of a sample delivery in 06/2013[3, 4]**

*The citation matching algorithm of CWTS*

The citation matching algorithm of CWTS was developed because it was observed that WoS fails to link a substantial number of cited references to cited articles. The distribution of these non-matching references or missed matches is extremely uneven. In some situations, the percentage of missed matches may be as high as 30% (Moed, 2002). According to Moed (2002), citation statistics at the level of individuals, institutions, journals, and countries could, therefore, be highly inaccurate and could be strongly affected by missed matches. The main aim of the citation matching algorithm of CWTS is to overcome the problem of missed matches as much as possible, while maintaining a high level of precision in the matching of citations. Because the citation matching algorithm of CWTS relies on bibliographic data provided by WoS, it is difficult to improve the precision of citation matches. As will be discussed in the subsection *Correct vs. incorrect matches in WoS*, most incorrect citation matches are caused by incorrect data in cited references in WoS, in particular by cited references that are, in fact, not listed in the citing article.

A very conservative approach to citation matching would be to use a strict rule that links a cited reference to a cited article only if all bibliographic fields of the cited reference match perfectly with the bibliographic fields of the cited article. Such a conservative approach would result in a substantial number of missed matches, as in the case of WoS. As explained earlier, this is because cited references often include inaccuracies. In order to minimize the problem of missed matches, a citation matching algorithm needs to be able to deal with inaccuracies in cited references. The citation matching algorithm of CWTS, therefore, relies not only on a single strict matching rule, but also on a series of less strict matching rules. These rules are applied iteratively in decreasing order of strictness.

The iterative, rule-based citation matching algorithm of CWTS works as follows. First, some preprocessing is applied to the bibliographic fields of both cited references and target articles. Numerical fields are cleaned by removing non-numeric characters, and textual fields are cleaned by removing diacritics and non-alphabetical characters. Furthermore, only the first initial of author names is retained. The citation matching algorithm then performs a series of matching attempts. Each iteration includes only those cited references that have not been matched to a cited article in one of the earlier iterations, and each iteration attempts to identify correct matches that have not been identified before while limiting the number of incorrect matches. The citation matching algorithm starts with the most restrictive matching rules (e.g., exact match on first author, publication year, publication name, volume number, starting page number, and DOI) and then proceeds with less restrictive matching rules (e.g., match on Soundex encoding[5] of the last name of the first author, publication year plus or minus one, volume number, and starting page number). The less restrictive matching rules allow for various types of inaccuracies in the bibliographic fields of cited references: fields with minor errors (e.g., publication year plus or minus one, or starting page number of the cited reference in-between the starting and end page numbers of the cited article), fields that have been interchanged (e.g., use of supplement number as volume number) and fields that have been omitted (e.g., match without taking into account the publication name). In all rules, the Levenshtein distance[6] is used to match the publication name of a cited reference to the publication name of a cited article. Finally, if, in a certain iteration, a cited reference can be matched to multiple cited articles, the cited article with the largest number of accumulated citations is selected.

### The citation matching algorithm of iFQ

The citation matching algorithm of iFQ has been developed as part of a project dedicated to error source analysis in bibliometrics. The algorithm was designed with the foremost intention to evaluate WoS's reference matching with regard to potential missed and incorrect matches and to enable large-scale analyses of their effects. It has not yet been used in a production environment. Like the CWTS algorithm, it can be described as an iterative, rule-based algorithm, starting from a very exact matching rule to rules with gradually increasing error-tolerance due to the permission of deviations or the omission of single or combinations of bibliographic metadata. The algorithm works with the following bibliographic fields: name and initials of the first author, abbreviated publication name[7], volume number, starting page number, publication year and DOI. Non-numeric characters are removed from volume and starting page numbers; non-alphanumeric characters are removed from author names and abbreviated publication names. The algorithm matches each cited reference indicating a specific year against every article of that year and additionally of the subsequent year. The iterative procedure is aborted for a cited reference-article pair if a match is achieved, but proceeds to check with regard to the whole data corpus in order to establish whether other matches can be achieved with a more exact or the same matching rule. It, therefore, allows non-unique matches of a single cited reference with several target articles. Since the algorithm has not been used in a production, no heuristic rule or procedure has been chosen to deal with these ambiguous matches. Only the match(es) with the most exact rule are stored. In the case of the textual metadata publication name and author name, the Damerau-Levenshtein distance metric[8] is applied. In the case of the numerical fields, volume number and starting page number, a numerical threshold for deviations is used and, in a reduced manner, a threshold for string deviations. All publication name abbreviations, which are delivered by WoS for journal records, are used for matching.

**Methodology**

In this section, the methodology applied in this research is described. It builds upon the study by Olensky (2015) and reuses its data corpus. The corpus consists of 300 target articles which were selected in a stratified, purposeful sampling process[9]. The strata represent a sub-universe of typical WoS[10] publications and cover the two science domains (natural sciences, and social sciences and humanities), six different disciplines (WoS subject category: Orthopedics, General & Internal Medicine, Multidisciplinary Chemistry, Sociology, Political Science and Education & Educational Research), two languages (English and German) and two (and three) publication years (1998 and 2003 as well as 2008 for the social sciences and humanities articles since the citation counts were rather low). For the 300 cited articles all matched cited references in the WoS Core Collection, consisting of the Science Citation Index Expanded (SCIE), the Social Sciences Citation Index (SSCI) and the Arts & Humanities Citation Index (A&HCI), were retrieved and also missed matches in the Cited Reference Search were identified for a citation window of 15 years (1998-2012). A total number of 3,968 cited references were detected.

For all 300 target articles and all source articles containing the 3,968 references, the bibliographic records in WoS were downloaded and the original articles (electronic version or a copy of the paper version) retrieved. It was not possible to obtain the original documents of 27 publications which were, therefore, excluded from the analysis. The original bibliographic data was manually extracted and the data values of the references recorded as-is (including any potential inaccuracies). Therefore, all references were manually screened to verify whether they really cited the target article indicated in WoS[11]. Additionally, the 300 target articles were queried in the databases of CWTS and iFQ and the matched cited references were retrieved. Only citations covered in all three databases were investigated[12]. Hence, a corpus of 3,975 cited references was analyzed to determine whether they were a correct match, a missed match or an incorrect match.

The last step of the research design encompassed a qualitative analysis of the missed matches in WoS (Figure 3), which were identified in the Cited Reference Search. In total, 244 cited references were found. The cited reference information of these references from WoS was extracted from the source articles' bibliographic data record and assessed against the cited reference information of correctly matched references. The data values of the following bibliographic fields were assessed: first author's last name, first and second initial, publication year, publication name, volume number and starting page number. The assessment process followed the methodology of Olensky (2015) and applied the Levenshtein distance function to identify discrepant values. Afterwards, they were manually assessed and annotated according to the taxonomy of bibliographic inaccuracies (Figure 1). In the present research the definition of Olensky (2015) is adopted which considers any non-conformity between the correct and the assessed value as an inaccuracy. The terms inaccuracy and discrepancy are used synonymously.

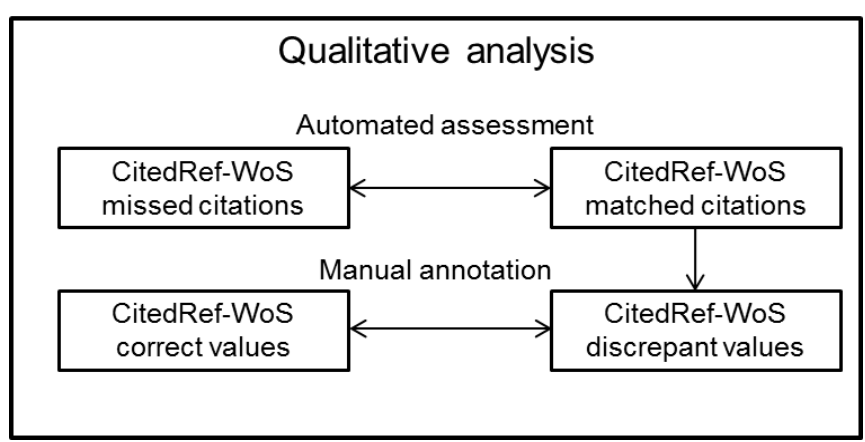

**Figure 3. Qualitative analysis of missed matches in WoS**

**Results**

*Quantitative results*

In this subsection, the results of the quantitative data analysis are presented, which are summarized in Table 1. We first look at the proportion of missed matches and the proportion of incorrect matches. We define the proportion of missed matches as the number of missed matches divided by the sum of the number of correct matches and the number of missed matches. Likewise, we define the proportion of incorrect matches as the number of incorrect matches divided by the sum of the number of correct matches and the number of incorrect matches. A proportion of missed matches of 6.19% were obtained for WoS and no incorrect matches were found. However, it should be emphasized that only *technically* correct and incorrect matches are considered here. It should be kept in mind that source articles were detected, that were linked to a target article through the cited reference information, even though the original source article did not cite the target article. This issue is discussed further in the subsection *Correct vs. incorrect matches in WoS*. Both CWTS and iFQ produced a much lower proportion of missed matches (CWTS: 1.34%; iFQ: 1.32%). The proportion of incorrect matches, however, is higher than in WoS (CWTS: 0.41%; iFQ: 0.71%). Interestingly, the records of the missed and the incorrect matches overlapped but were not completely identical for CWTS and iFQ. Our results can also be summarized in terms of the well-known concepts of precision and recall. We define precision and recall as, respectively,

$$\text{Precision} = \frac{\text{Correct matches}}{\text{Correct matches} + \text{Incorrect matches}},$$

and

$$\text{Recall} = \frac{\text{Correct matches}}{\text{Correct matches} + \text{Missed matches}}.$$

Hence, precision equals one minus the proportion of incorrect matches, while recall equals one minus the proportion of missed matches. A frequently used measure in which precision and recall are combined is the $F_1$ score, defined as the harmonic mean of precision and recall. The $F_1$ score equals

$$F_1 = 2 \times \frac{\text{precision} \times \text{recall}}{\text{precision} + \text{recall}}.$$

Table 1 presents precision, recall and $F_1$ score for all three data sources. In terms of $F_1$ score, CWTS performs best, closely followed by iFQ. WoS achieves the lowest $F_1$ score, even though this score is still close to 97%.

**Table 1. Correct, incorrect and missed matches in the three data sources and the resulting precision, recall and $F_1$ scores**

|  | WoS | | CWTS | | iFQ | |
| --- | --- | --- | --- | --- | --- | --- |
|  | # | % | # | % | # | % |
| Correct matches | 3,697 | 93.81 | 3,888 | 98.66 | 3,889 | 98.68 |
| Incorrect matches | 0 | 0 | 16 | 0.41 | 28 | 0.71 |
| Missed matches | 244 | 6.19 | 53 | 1.34 | 52 | 1.32 |
| Precision | | 100.00 | | 99.59 | | 99.29 |
| Recall | | 93.81 | | 98.66 | | 98.68 |
| **$F_1$ score** | | **96.81** | | **99.12** | | **98.98** |

*Qualitative results*

In this subsection, the results of the qualitative data analysis of the 244 missed matches of WoS, the 53 missed matches of CWTS, and the 52 missed matches of iFQ[13] are presented. The missed matches of CWTS and iFQ are also missed by WoS, hence, a total of 244 cited references were investigated[14]. The distribution of inaccuracies according to the taxonomy of bibliographic inaccuracies is summarized in Figure 4. Each subcategory consists of a different number of IACs. In order to compare the number of inaccuracies in each subcategory (e.g. *Added data values*, *Disarranged data values*, etc.), they were normalized by the number of IACs present in each subcategory, since each consists of a different number of IACs (cf. Figure 1). This facilitates a comparability of the shares of inaccuracy subcategories by source data values for the three data sources.

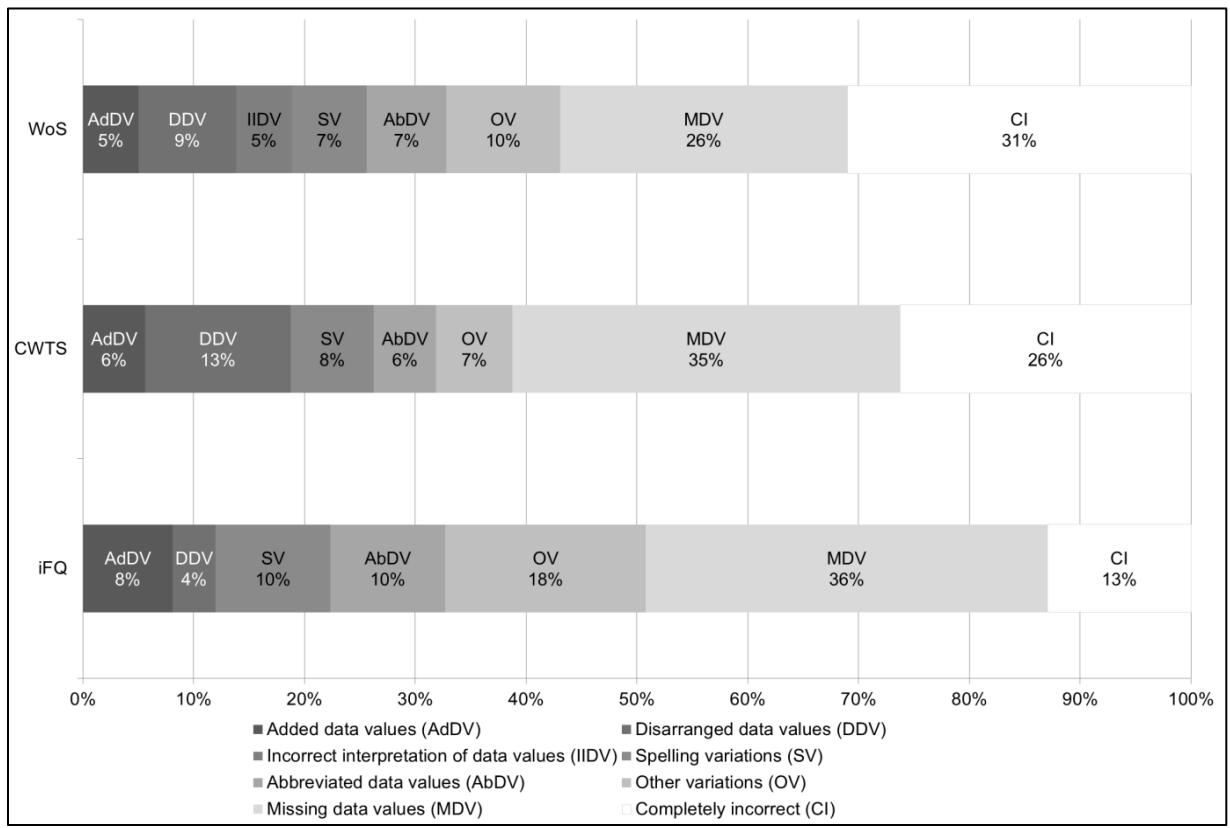

**Figure 4. Comparison of inaccuracy subcategories in missed citations for each data source**

In general, it can be observed that the CWTS and the WoS data show two distinct peaks (*Missing data values* and *Completely incorrect*), while the remaining categories are quite evenly distributed. iFQ, on the other hand, shows one larger peak in the category *Missing data values*, followed by two smaller peaks (*Other variations* and *Completely incorrect*), and the other remaining four are also fairly evenly distributed. Hence, all three data sources indicate that *Missing data values* and *Completely incorrect* data values represent major problems in cited references that were missed in the matching process.

In the category *Added data values*, iFQ has a slightly higher share than the other two data sources (8% vs. 6% and 5%). However, in the category *Disarranged data values* iFQ performs better than the other two sources, with only half the share compared to WoS and only a third compared to CWTS. Inaccuracies pertaining to the category *Incorrect interpretation of data values* are only present in cited references missed by WoS, i.e. the algorithms of CWTS and iFQ can handle these inaccuracies well. The shares of values containing *Spelling variations* are almost equally high in all three data sources and range between 7% and 10% respectively. The lowest share of *Abbreviated data values* was 6% for the CWTS data; iFQ and WoS have slightly higher shares of 7% and 10%. iFQ has by far the largest share of *Other variations* compared to the other data sources, which means that inaccuracies in this category cause more problems for the iFQ algorithm. *Missing data values* are the most problematic for the citation matching algorithms of CWTS and iFQ, while *Completely incorrect* data values pose the biggest problem for the WoS algorithm. This category is also prominently represented in the missed citations of CWTS and with a smaller share in the iFQ data.

Comparing the citation matching algorithms of all three data sources on the level of the single inaccuracy codes (Figure 5) enables one to determine what types of inaccuracies the citation matching algorithms of CWTS and iFQ are able to handle compared to WoS. The IACs on the horizontal axis are sorted by their frequency in WoS in absolute numbers of occurrence. The distribution of inaccuracies is slightly different for the three data sources. While the most frequent four types of inaccuracies (*D Completely incorrect*, *E Omitted*, *G Interchanged fields* and *T Plus/Minus*) are the same for WoS and CWTS, but differently ranked, iFQ only shares the IACs *T Plus/Minus* and *E Omitted* in its top four inaccuracy types. For iFQ, two other frequently recurring IACs are *R Punctuation* and *I Abbreviation*.

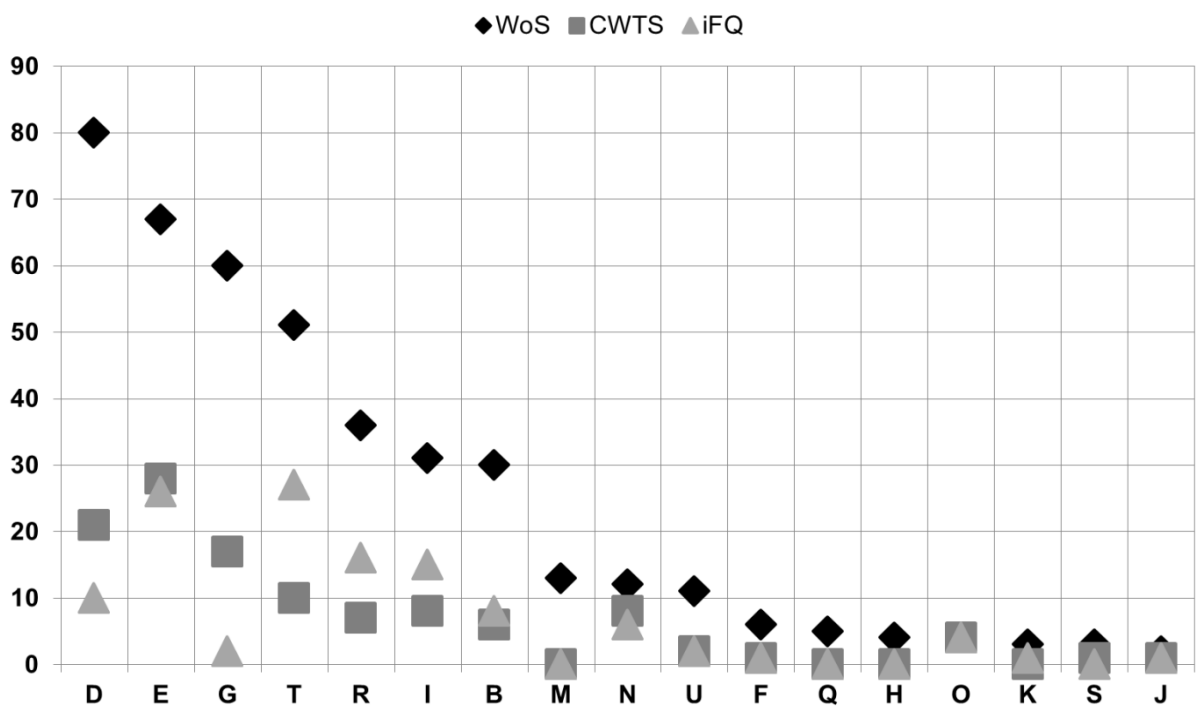

**Figure 5. Types of inaccuracies occurring in the cited reference information of missed citations – WoS, CWTS, iFQ**

The biggest problems in missed citations in WoS are the IACs *D Completely incorrect*, *E Omitted*, *G Interchanged fields* and *T Plus/Minus* with a total number of occurrences between 80 and 51. A second group of inaccuracies ranging between 36 and 30 occurrences can be observed, and the remaining inaccuracies lie between 13 and 2 incidences. In the CWTS data, the most frequent inaccuracy type is *E Omitted* (28 occurrences), followed by *D Completely incorrect and G Interchanged fields* with a total number of occurrences of 21 and 17 respectively. The rest of the inaccuracies occur 10 times or less. For the algorithm of iFQ, the inaccuracies *T Plus/Minus* and *E Omitted* present the greatest challenges with a total number of occurrences of 27 and 26. Another group of three inaccuracy types (*R Punctuation*, *I Abbreviation* and *D Completely incorrect*) ranges between 16 and 10 occurrences. The inaccuracies *M Incorrect interpretation of author names*, *H Jumbled value* and *Q Special character* do not occur at all in either dataset, hence the algorithms of CWTS and iFQ can handle them very well.

In WoS, 44% of the cited references were not matched to their target article because of a single inaccuracy in the cited reference information. In the CWTS data, the figure was 28% and in the iFQ data 15%. Hence, in the iFQ database a missed match is more likely to be caused by a combination of multiple inaccuracies rather than just a single inaccuracy.

## Discussion

*Correct vs. incorrect matches in WoS*

As mentioned in the *Methodology* section, the manual verification of the cited references in the source articles allowed us to determine whether all *technically* correct matches were also *empirically* correct matches. The intellectual verification of the results revealed 29 incorrect matches, i.e. false positive matches, in our WoS data corpus. These were source articles that had been linked by incorrect cited reference information to a target article, which was not cited in the original source article[15]. A closer investigation of the incorrect matches in WoS revealed that they were caused by an inaccurate data extraction process by WoS. They were also matched in the other two data sources because of the perfectly correct cited reference information, which caused a *technically* correct, but *empirically* incorrect match.

25 of these cited references were matched to the cited reference information of an incorrect target article by WoS due to incorrect matching of the extracted cited reference information: The first four letters of the first author's last name, as well as either the volume number and the publication year or the volume number and the starting page number of the original reference in the source article matched the cited reference information of the target article to which they were matched. Apparently neither the initials nor the publication name were considered at all. Two of the cited references were matched to the cited reference information of an incorrect target article because they had the same first author, publication name, volume number and pagination in the original reference of the source article. However, the other authors, the publication year and the article title did not match the target article. Hence, the incorrect cited reference information for these two cited references was partly caused by the citing authors who had cited an incorrect volume number and pagination. For the remaining two citations, their original publications did not contain any references at all; therefore, it is not possible to explain why the incorrect cited reference information was extracted. The following examples illustrate the two types of incorrect matches (no example is given for the source articles where no references were found).

The following reference from the source article[16]

> *Holland, S.J., Peles, E., Pawson, T., and Schlessinger, J. (1998). Cell-contact-dependent signaling in axon growth and guidance: Ephreceptor tyrosine kinases and receptor protein tyrosine phosphatase beta. Curr. Opin. Neurobiol. 8, 117–127.*

was converted into the cited reference information (matching the first four letters of the first author's last name, volume number and the publication year):

```
Hollstein B, 1998, BERL J SOZIOL, V8, P7
```

The following reference from the source article[17]

> *Schannwell CM, Schoebel FC, Zimmermann T et al. (2000) Linksventrikuläre diastolische Funktion in der normalen Schwangerschaft. Dtsch Med Wochenschr 123:957–964*[18]

was converted into the cited reference information:

```
Schannwell CM, 1998, DEUT MED WOCHENSCHR, V123, P957, DOI
10.1055/s-2007-1024104
```

Considering only *empirically* correct matches, the $F_1$ score from the quantitative results has to be slightly adjusted downwards (Table 2). Additionally, the matching algorithm of iFQ allows for ambiguous matches. Eight of the incorrect matches in the iFQ data are actually ambiguous matches, i.e. they were matched to the correct target article as well as to another incorrect target article. Considering this aspect and assuming that all ambiguous matches were verified manually, the $F_1$ score could be further corrected from 98.61% to 98.71% for the *empirically* correct matches, bringing the result closer to that of CWTS and reducing the delta to 0.03%. However, it should be emphasized that this describes a best case scenario for iFQ. iFQ has not yet decided how to handle such ambiguous matches automatically and manual verification is quite difficult to implement.

**Table 2. F₁ score – revisited (in %)**

| F₁ score | WoS | CWTS | iFQ |
|---|---|---|---|
| Technically correct matches | 96.81 | 99.12 | 98.98 |
| Empirically correct matches | 96.41 | 98.74 | 98.61 |

In the *technically* correct matches, 4 correct matches stemmed from double records[19] and 2 from source articles which were corrections of the target article in the WoS data. The double records were excluded from the corpus of the *empirically* correct matches. WoS links corrections (not all, but most of them) to their original articles via the cited reference information. Hence, technically speaking, these 2 matches are correct and are counted as such. However, in the light of a citation analysis, they do not constitute a valid citation and would have to be excluded as well.

*Interpretation of the differences in the matching results of CWTS and iFQ*

As shown in Figure 5, the citation matching algorithms of CWTS and iFQ are both able to handle inaccuracies in cited references much better than the citation algorithm of WoS. However, there are also differences in the types of inaccuracies to which the CWTS and iFQ matching algorithms are prone. The IACs that show the greatest differences are *G Interchanged fields*, *T Plus/minus*, *D Completely incorrect*, *R Punctuation*, and *I Abbreviation*.

The IAC *G Interchanged fields* is better handled by the iFQ algorithm than by that of CWTS. Most of the cited references' starting page numbers or volume numbers have been interchanged with the issue numbers. The inaccuracies are not corrected by the iFQ algorithm, but the data values are treated as if they were *Completely incorrect* (IAC *D*). In a few cases, starting page numbers are interchanged with volume numbers and vice versa, which are matched by iFQ with specific matching rules for these scenarios. Hence, the main reason why the iFQ missed matches contain fewer IACs *G Interchanged fields* than the CWTS data is the generally higher tolerance of incorrect data values especially with regard to numeric data.

The IAC *T Plus/Minus*, on the contrary, poses a greater problem for the iFQ algorithm than for the CWTS algorithm. The iFQ algorithm only allows the publication year of the target article to be one year later than that of the cited reference, while all other bibliographic fields, including the DOI, have to match exactly. Hence, iFQ's rather strict handling of publication years causes most missed matches in its citation matching. In the case of the CWTS algorithm, the publication year of the target article is allowed to be either one year later or one year earlier than that of the cited reference and a perfect match on DOI and other fields is not required. The less strict approach of the CWTS algorithm explains its superior performance.

The IAC *D Completely incorrect* is better handled by the iFQ algorithm than by the CWTS algorithm. Similar to the IAC *G Interchanged fields*, this is due to the generally higher error tolerance of the iFQ algorithm. However, completely incorrect data values pose a problem for both algorithms. In particular when they occur in more than one bibliographic field, they are only rectifiable at the cost of precision.

The IAC *R Punctuation* occurs more often in the iFQ than in the CWTS missed matches. However, punctuation errors as such neither pose any difficulties for CWTS nor iFQ. Missed citations are actually caused by the co-occurring inaccuracies, predominantly by inaccuracies in publication years. In contrast, one reference was found in the WoS data that was not matched due to a difference in the punctuation of the first initial.

Cited references with the IAC *I Abbreviation* missed by the iFQ algorithm, but matched by the CWTS algorithm, are mostly caused by different publication name abbreviations which do not match the abbreviations provided by WoS and were not managed by the threshold determined in the Damerau-Levenshtein function either, sometimes in combination with other inaccuracies. In the CWTS algorithm, the publication name abbreviation of a cited reference is not matched with the publication name abbreviation of a target article, but with the full publication name of a target article, and a sophisticated fuzzy matching approach is adopted. This approach uses the Levenshtein distance with a variable threshold. Hence, the CWTS approach seems to work slightly better for abbreviated publication names than that of iFQ.

The citation matching algorithms of CWTS and iFQ perform fairly similarly with regard to all other IACs. Of all other IACs, most problems are caused by IAC *E Omitted*. Cited references with IAC *E Omitted*, which are missed matches, usually contain combinations of multiple missing or missing and incorrect values. In most cases these combinations exceed the error-tolerance of both algorithms.

## Conclusion

This research investigated the performance of the citation matching algorithms of CWTS and iFQ in comparison to WoS. It was found that the algorithms of the bibliometric research groups perform better than the WoS algorithm. Apparently the WoS algorithm does not allow for any variations, since the matches are all *technically* correct and only produce missed matches due to inaccuracies. However, the manual verification of the *technically* correct matches also revealed incorrect matches: the cited reference information in source articles can also contain completely incorrect links to target articles which were not cited by the source articles, caused by incorrect data extraction and/or matching between the original reference and the cited reference information, respectively. Because such issues can only be identified manually, it is not possible to determine the extent of this problem for the entire WoS. In our corpus it affected 0.79% of the references, i.e. 29 citations, and of these 31% could have been avoided by an additional rule that checks whether the domain, i.e. natural sciences or social sciences, of the source and the target article match and, therefore, whether they are likely to be linked. Ideally, WoS would already use this rule in its citation extraction process. The incorrect WoS matches naturally have repercussions on the citation matching algorithms of CWTS and iFQ, since they can only rely on the information provided by WoS. Given this limitation, the algorithms of CWTS and iFQ perform excellently and seem to have found a good balance between precision and recall in their citation matching.

In total, the CWTS algorithm missed one citation more than that of iFQ. However, the iFQ algorithm also produced more incorrect matches. Furthermore, the two algorithms do not find the same correct matches. The iFQ algorithm finds correct matches which are not detected by the CWTS algorithm on account of its relatively error-tolerant matching rules. However, the more error-tolerant matching rules also produce incorrect matches which are not simple to surmount without the full reference data, including article titles. Hence, it seems appropriate to modify the error-tolerance. The relatively exact handling of publication years by the iFQ algorithm, which is responsible for missed citations that were, however, matched by the CWTS algorithm, could be adjusted accordingly. The results indicate that the CWTS algorithm could be improved by including more matching rules capable of handling interchanged fields better (e.g. a matching rule for cases where the starting page number and volume number of a cited reference have been interchanged). It was also found that, in some cases, it is not necessarily a certain type of inaccuracy that causes a citation to be missed in the matching process, but rather a question of the weighting and the inclusion or exclusion of specific bibliographic fields in the algorithm which determines whether a match is correct or missed. Overall, the findings corroborate Olensky's conclusion (2015) that, if WoS were to provide the full cited reference information, like Scopus, the matching algorithms could take additional authors as well as the article title into account, which would open up new opportunities for variation thresholds in the citation matching process.

---

[1] Cf. Table A1 in the Appendix for the complete list of inaccuracy codes used in the present research.

[2] Our requests to obtain information from Thomson Reuters on its data ingestion processes remained unanswered.

[3] The WoS UT of this source article is 000319361500023; the WoS UT of this target article is: 000315372400005.

[4] The WoS UT of this source article is 000319410800003; the PubMed ID of this target article is: 23278144.

[5] Soundex is a phonetic algorithm for indexing names by sound, as pronounced in English. It aims to encode homophones to the same representation so that they can be matched even in the case of minor differences in spelling (Knuth, 1997).

[6] The Levenshtein distance is a string metric for measuring the difference between two strings. It indicates the distance between two strings as the minimum number of single-character edits (i.e., insertions, deletions, and substitutions) required to change one string into the other.

[7] Or the complete publication name if there is no standard abbreviation for the publication name.

[8] The Damerau-Levenshtein distance extends the original Levenshtein distance function by including the transposition of two adjacent characters in its allowable operations.

[9] For a more detailed description of the data selection process, cf. Olensky (2015). A purposeful stratified sample was used in this previous study because its main goal was to investigate the types of inaccuracies occurring in bibliographic references and organize them into a comprehensive taxonomy of bibliographic inaccuracies. Reusing the data corpus, on the one hand, allowed us to build on a manually verified corpus; on the other hand, it also facilitated a qualitative in-depth analysis of how the citation matching algorithms of the bibliometric research groups handle this wide range of inaccuracy types.

[10] The web access of the WoS Core Collection was used for this study as available via the Humboldt-Universität zu Berlin.

[11] This was accomplished by verifying all available bibliographic metadata. In the majority of cases a matching article title was the deciding evidence.

[12] There are slight discrepancies between the three databases in the coverage of the publications used in this analysis and, therefore, only citations included in all three databases are taken into account.

[13] In some targeted checks, we detected cited reference information in the CWTS and iFQ databases that differed from that available as a download from the WoS web interface, which points to further inconsistencies in the WoS data. The three references are listed in the Appendix in Table A2.

[14] References that were missed in all three databases are listed in the Appendix in Table A3.

[15] A full list of the incorrect matches caused by incorrect data extraction and/or incorrect cited reference information in WoS are given in the Appendix in Table A4.

[16] The WoS UT of this source article is: 000084485900003.

[17] The WoS UT of this source article is: 000181820100013.

[18] The correct volume number, issue number and pagination for this publication should have been 125(37): 1069-1073; DOI 10.1055/s-2000-7356

[19] A list of the double records is given in the Appendix in Table A5.

## References

Adriaanse, L. S., & Rensleigh, C. (2013). Web of Science, Scopus and Google Scholar. A content comprehensiveness comparison. *The Electronic Library, 31*(6), 727–744. http://dx.doi.org/10.1108/el-12-2011-0174

Archambault, É., Campbell, D., Gingras, Y., & Larivière, V. (2009). Comparing bibliometric statistics obtained from the Web of Science and Scopus. *Journal of the American Society for Information Science and Technology, 60*(7), 1320–1326. http://dx.doi.org/10.1002/asi.21062

Chang, C.-L., McAleer, M., & Oxley, L. (2011). Great expectatrics: Great papers, great journals, great econometrics. *Econometric Reviews, 30*(6), 583–619. http://dx.doi.org/10.1080/07474938.2011.586614

Franceschini, F., Maisano, D., & Mastrogiacomo, L. (2013a). A novel approach for estimating the omitted-citation rate of bibliometric databases with an application to the field of bibliometrics. *Journal of the American Society for Information Science and Technology, 64*(10), 2149–2156. http://dx.doi.org/10.1002/asi.22898

Franceschini, F., Maisano, D., & Mastrogiacomo, L. (2013b). Research quality evaluation: comparing citation counts considering bibliometric database errors. *Quality & Quantity*, 1–11. http://dx.doi.org/10.1007/s11135-013-9979-1

Galvez, C., & de Moya-Anegón, F. (2006). The unification of institutional addresses applying parametrized finite-state graphs (P-FSG). *Scientometrics, 69*(2), 323–345. http://dx.doi.org/10.1007/s11192-006-0156-3

Galvez, C., & de Moya-Anegón, F. (2007). Standardizing formats of corporate source data. *Scientometrics, 70*(1), 3–26. http://dx.doi.org/10.1007/s11192-007-0101-0

Garfield, E. (1990). Journal editors awaken to the impact of citation errors. *Essays of an Information Scientist, 13*(41), 367. Retrieved September 24, 2014 from http://garfield.library.upenn.edu/volume13.html

Hildebrandt, A. L., & Larsen, B. (2008). *Reference and citation errors: A study of three law journals*. Presented at the 13th Nordic Workshop on Bibliometrics and Research Policy. 11-12 September 2008, Tampere, Finland.

Jacsó, P. (2004). The future of citation indexing: An interview with Eugene Garfield. *Online, 28*(1), 38–40.

Knuth, D. E. (1997). *The art of computer programming Volume 3*. Reading, Massachusetts: Addison-Wesley.

Larsen, B., Hytteballe Ibanez, K., & Bolling, P. (2007). *Error rates and error types for the Web of Science algorithm for automatic identification of citations*. Presented at the 12th Nordic Workshop on Bibliometrics and Research Policy. 13-14 September 2007, Copenhagen, Denmark.

Liang, L., Zhong, Z., & Rousseau, R. (2014). Scientists' referencing (mis)behavior revealed by the dissemination network of referencing errors. *Scientometrics*. http://dx.doi.org/10.1007/s11192-014-1275-x

Meho, L. I., & Rogers, Y. (2008). Citation counting, citation ranking, and *h*-index of human-computer interaction researchers: A comparison of Scopus and Web of Science. *Journal of the American Society for Information Science and Technology, 59*(11), 1711–1726. http://dx.doi.org/10.1002/asi.20874

Meho, L. I., & Yang, K. (2007). Impact of data sources on citation counts and rankings of LIS faculty: Web of Science vs. Scopus and Google Scholar. *Journal of the American Society for Information Science and Technology, 58*(13), 2105–2125. http://dx.doi.org/10.1002/asi.20677

Meyer, C. A. (2008). Reference accuracy: Best practices for making the links. *Journal of Electronic Publishing, 11*(2). http://dx.doi.org/10.3998/3336451.0011.206

Moed, H. F. (2002). The impact-factors debate: the ISI's uses and limits. *Nature, 415*(6873), 731–732. http://dx.doi.org/10.1038/415731a

Moed, H. F. (2005). *Citation analysis in research evaluation. Information Science and Knowledge Management: Vol. 9*. Dordrecht: Springer. http://dx.doi.org/10.1007/1-4020-3714-7

Moed, H. F., & Vriens, M. (1989). Possible inaccuracies occurring in citation analysis. *Journal of Information Science, 15*(2), 95-107. http://dx.doi.org/10.1177/016555158901500205

Neuhaus, C., & Daniel, H.-D. (2008). Data sources for performing citation analysis: An overview. *Journal of Documentation, 64*(2), 193–210. http://dx.doi.org/10.1108/00220410810858010

Olensky, M. (2012). How is bibliographic data accuracy assessed? In É. Archambault, Y. Gingras, & V. Larivière (Eds.), *Proceedings of the 17th International Conference on Science and Technology Indicators* (pp. 628–639). Montréal, Canada. Retrieved September 24, 2014 from http://2012.sticonference.org/index.php?page=proc

Olensky, M. (2015). Data accuracy in bibliometric data sources and its impact on citation matching. Doctoral dissertation. Humboldt-Universität zu Berlin (Germany). Retrieved January 26, 2015 from http://edoc.hu-berlin.de/dissertationen/olensky-marlies-2014-12-17/PDF/olensky.pdf

Schmidt, M. (2012). Development and evaluation of a match key for linking references to cited articles In É. Archambault, Y. Gingras, & V. Larivière (Eds.), *Proceedings of the 17th International Conference on Science and Technology Indicators* (pp. 707–718). Montréal, Canada. Retrieved September 24, 2014 from http://2012.sticonference.org/index.php?page=proc

Sweetland, J. H. (1989). Errors in bibliographic citations: A continuing problem. *The library quarterly, 59*(4), 291–304. http://dx.doi.org/10.1086/602160

Tunger, D., Haustein, S., Ruppert, L., Luca, G., & Unterhalt, S. (2010). "The Delphic Oracle": An analysis of potential error sources in bibliographic databases. In CWTS (Ed.), *Proceedings of the 11th International Conference on Science and Technology Indicators* (pp. 282–283). Leiden, Netherlands.

van Raan, A. F. J. (2005). Fatal attraction: Conceptual and methodological problems in the ranking of universities by bibliometric methods. *Scientometrics, 62*(1), 133–143. http://dx.doi.org/10.1007/s11192-005-0008-6

# Appendix

**Table A1. List of inaccuracy codes occurring in the qualitative assessment of missed matches (Olensky, 2015)**

| Inaccuracy code | Name | Examples | |
|---|---|---|---|
| | | **Correct value** | **Incorrect value** |
| B | Spelling error | Arduengo | Aduengo |
| D | Completely incorrect | Journal of Curriculum Studies | Studies in Higher Education |
| E | Omitted | Pant, HA | Pant, H |
| F | Cropped | P827 | P82 |
| G | Interchanged fields | V37, P52 | V52, P1 |
| | G1 holds issue no | | |
| | G2 holds starting page number | | |
| | G3 holds ending page number | | |
| | G4 holds volume no | | |
| | G5 holds last name | | |
| | G6 holds first initial | | |
| | G7 holds second initial | | |
| H | Jumbled value | P654 | P564 |
| I | Abbreviation | Chem unserer Zeit | Chem Z |
| J | Partially incorrect | Giessler | GoetzGiessler |
| K | Space | De Castell | DeCastell |
| M | Incorrect interpretation of author names | Garcia-Elias M | Elias MG |
| N | Additional information | Deut Med Wochenschr | In press Deut Med Wochenschr |
| O | Incorrect order of authors | First author: Raghunathan, R Second author: Shanmugasundara M | First author: Shanmugasundara M Second author: Raghunathan, R |
| Q | Special character | Köster | Koster |
| R | Punctuation | Vobruba G. | Vobruba G |
| S | Padded | V30 | V300 |
| T | Plus/Minus (denotes a data value that is correct if the number 1 or 2 is added to, or subtracted from, the data value. The calculation can either be made on the total number or just on one of the digits) | P251 | P261 |
| U | Full first name | Brauninger Thomas | Brauninger T |

**Table A2. List of references with differing cited reference information in CWTS, iFQ and WoS**

| WoS_UT_Target | WoS_UT_Source | Cited reference information – CWTS, iFQ | Cited reference information – WoS |
|---|---|---|---|
| 000184794300022 | 000258275100013 | ALTENMUELLER E, 2003, HAND CLIN, V19, P1 | ALTENMULLER E, 2003, HAND CLIN, V19, P1 |
| 000184794300022 | 000229736400004 | ALTENMUELLER E, 2003, HAND CLIN, V19, P1 | ALTENMULLER E, 2003, HAND CLIN, V19, P1 |
| 000073915400012 | 000275640700014 | Heimcke J., 1998, HETEROATOM CHEM, P439 | Heinicke J., 1998, HETEROATOM CHEM, V9, P3 |

**Table A3. Cited reference information as provided in the download function for web interface users by WoS – missed matches in all three data sources**

| WoS_UT_Target | WoS_UT_Source | Cited reference information – missed match | Cited reference information – correct match |
|---|---|---|---|
| 000182480200014 | 000230648100018 | SHI DQ, 2005, HETEROATOM CHEM, V14, P266 | Shi DQ, 2003, HETEROATOM CHEM, V14, P266, DOI 10.1002/hc.10139 |
| 000073141300009 | 000247620200051 | SHANMUGASUNDARA.M, 1998, HETEROATOM CHEM, P327 | Raghunathan R, 1998, HETEROATOM CHEM, V9, P327, DOI 10.1002/(SICI)1098-1071(1998)9:3<327::AID-HC9>3.0.CO;2-6 |
| 000184794300022 | 000258275100013 | ALTENMULLER E, 2003, HAND CLIN, V19, P1 | Altenmuller E, 2003, HAND CLIN, V19, P523, DOI 10.1016/S0749-0712(03)00043-X |
| 000184794300022 | 000229736400004 | ALTENMULLER E, 2003, HAND CLIN, V19, P1 | Altenmuller E, 2003, HAND CLIN, V19, P523, DOI 10.1016/S0749-0712(03)00043-X |
| 000181956900002 | 000245585800010 | NEUMEISTER MW, HAND CLIN FEB, V19, P1 | Neumeister MW, 2003, HAND CLIN, V19, P1, DOI 10.1016/S0749-0712(02)00141-5 |
| 000073670600011 | 000089361500016 | LICHTMAN DM, 1988, HAND CLIN, V14, P265 | Lichtman DM, 1998, HAND CLIN, V14, P265 |
| 000253418400006 | 000300625200004 | Van Driel J.H., 2006, J CURRICULUM STUD, V40, P107 | Van Driel JH, 2008, J CURRICULUM STUD, V40, P107, DOI 10.1080/00220270601078259 |
| 000073436100002 | 000236249200009 | LENSMIRE T, 1998, J CURRICULUM STUD, V30, P251 | Lensmire TJ, 1998, J CURRICULUM STUD, V30, P261, DOI 10.1080/002202798183611 |
| 000073436100002 | 000175953600005 | LENSMIRE T, 1998, J CURRICULUM STUDIES | Lensmire TJ, 1998, J CURRICULUM STUD, V30, P261, DOI 10.1080/002202798183611 |
| 000083306800004 | 000232270000003 | PHILLIPSHOWARD PA, 1999, J TRAVEL MED, V77, P141 | Phillips-Howard PA, 1998, J TRAVEL MED, V5, P121, DOI 10.1111/j.1708-8305.1998.tb00484.x |
| 000181590500001 | 000318923300004 | PAGDEN, 2003, POLIT THEORY, V2, P171 | Pagden A, 2003, POLIT THEORY, V31, P171, DOI 10.1177/0090590702251008 |
| 000186429500003 | 000208330900003 | Nasstrom Sofia, 2004, POLIT THEORY, V31, P818 | Nasstrom S, 2003, POLIT THEORY, V31, P808, DOI 10.1177/0090591703252158 |

| 000252591500001 | 000290473600001 | MARKELL, 2008, POLIT THEORY, V36, P12 | Markell P, 2008, POLIT THEORY, V36, P9, DOI 10.1177/0090591707310084 |
| 000258077500002 | 000257000200002 | ROOVER J, 2008, POLITICAL THEORY | De Roover J, 2008, POLIT THEORY, V36, P523, DOI 10.1177/0090591708317969 |
| 000186430200005 | 000314511000003 | Freitag Markus, 2010, POLIT VIERTELJAHR, V44, P348 | Freitag M, 2003, POLIT VIERTELJAHR, V44, P348, DOI 10.1007/s11615-003-0068-2 |
| 000187426800003 | 000227826600009 | HOFMANN T, 2003, CHEM UNSERER Z, V6, P388 | Schieberle P, 2003, CHEM UNSERER ZEIT, V37, P388, DOI 10.1002/ciuz.200300305 |
| 000072551300002 | 000071393800006 | ARDUENGO AJ, IN PRESS CHEM UNSERE | Arduengo AJ, 1998, CHEM UNSERER ZEIT, V32, P6, DOI 10.1002/ciuz.19980320103 |
| 000077273300003 | 000186395900007 | STRAHO S, 2002, DEUT MED WOCHENSCHR, V123, P1410 | Strahl S, 1998, DEUT MED WOCHENSCHR, V123, P1410, DOI 10.1055/s-2007-1024196 |
| 000071599100001 | 000253088400011 | BERGANT AM, 1981, DEUT MED WOCHENSCHR, V23, P35 | Bergant AM, 1998, DEUT MED WOCHENSCHR, V123, P35, DOI 10.1055/s-2007-1023895 |
| 000075466100001 | 000080889900005 | SCHANNWELL CM, 1998, IN PRESS DTSCH MED W | Schannwell CM, 1998, DEUT MED WOCHENSCHR, V123, P957, DOI 10.1055/s-2007-1024104 |
| 000073461100001 | 000073461900010 | MESSMANN H, IN PRESS DTSCH MED W | Messmann H, 1998, DEUT MED WOCHENSCHR, V123, P515, DOI 10.1055/s-2007-1024003 |
| 000072512500004 | 000252288600014 | RUEGER JM, 1998, Knochenersatzmittel, V27, P71 | Rueger JM, 1998, ORTHOPADE, V27, P72, DOI 10.1007/PL00003481 |
| 000261119900005 | 000259460500007 | PANT HA, Z PADAGOGIK IN PRESS | Pant HA, 2008, Z PADAGOGIK, V54, P827 |

**Table A4. List of the incorrect matches caused by incorrect data extraction and incorrect cited reference information in WoS**

| WoS_UT_Target | WoS_UT_Source | Cited reference information – WoS | Original reference |
|---|---|---|---|
| 000073119800002 | 000181755500018 | Hollstein B, 1998, BERL J SOZIOL, V8, P7 | Hollander E (1998) Treatment of obsessive-compulsive spectrum disorders with SSRIs. Br J Psychiatry 8 (Suppl): 7–12 |
| 000073119800002 | 000084485900003 | Hollstein B, 1998, BERL J SOZIOL, V8, P7 | Holland, S.J., Peles, E., Pawson, T., and Schlessinger, J. (1998). Cell-contact-dependent signaling in axon growth and guidance: Ephreceptor tyrosine kinases and receptor protein tyrosine phosphatase beta. Curr. Opin. Neurobiol. 8, 117–127. |
| 000076443500007 | 000228255900003 | Offe C, 1998, BERL J SOZIOL, V8, P359 | Offenbacher S, Farr DH, Goodson JM. Measurement of prostagiandin E in crevicular fluid, J Clin Periodontol 1981; 8:359-367 |
| 000076443500007 | 000174834200012 | Offe C, 1998, BERL J SOZIOL, V8, P359 | Offenbacher, S., Farr, D.H., Goodson, J.M., 1981. Measurement of prostaglandin E in crevicular fluid. J. Clin. Periodont. 8, 359–367. |
| 000076443500010 | 000167374800009 | Albrow M, 1998, BERL J SOZIOL, V8, P411 | Thomas W. Albrecht & Sarah J. Smith. Corporate Loan Securization: Selected Legal and Regulatory Issues, 8 DUKE J. COMP. & INT'L L. 411, 414 (1998) |
| 000076443500010 | 000090094400001 | Albrow M, 1998, BERL J SOZIOL, V8, P411 | Thomas W . Albrecht & Sarah J . Smith, Corporate Loan Securitization: Selected Legal and Regulatory Issues, 8 DUKE J. COMP. & INT'L L. 411, 433 & nn.90-93 (1998) |
| 000076443500010 | 000087564800012 | Albrow M, 1998, BERL J SOZIOL, V8, P411 | Thomas W. Albrecht & Sarah J. Smith. Corporate Loan Securization: Selected Legal and Regulatory Issues, 8 DUKE J. COMP. & INT'L L. 411, 414 (1998) |
| 000074825600004 | 000084295000021 | Willems H, 1998, BERL J SOZIOL, V8, P201 | Williams, A. E. and Bradley, T. J. (1998). The effect of respiratory pattern on water loss in desiccation-resistant Drosophila melanogaster. J. Exp. Biol. 201, 2953–2959. |
| 000072590900002 | 000222356900009 | James MA, 1998, HAND CLIN, V14, P1 | James, W. 1998. 'One of us': Marcel Mauss and 'English' anthropology. In Marcel Mauss: a centenary tribute (eds) W. James & N. Allen, 1-26. Oxford: Berghahn. |
| 000072590900002 | 000224959600001 | James MA, 1998, HAND CLIN, V14, P1 | Jameson, M.L. (1998) Phylogenetic analysis of the subtribe Rutelina and revision of the Rutela generic groups (Coleoptera: Scarabaeidae: Rutelinae: Rutelini). Bulletin of the University of Nebraska State Museum, 14 (1997), 1-184. |
| 000072590900002 | 000222091700004 | James MA, 1998, HAND CLIN, V14, P1 | Jameson, M. L. (1998). Phylogenetic analysis of the subtribe Rutelina and revision of the Rutela generic groups (Coleoptera: carabaeidae: Rutelinae: Rutelini). Bulletin of the University of Nebraska State Museum 14, 1–184. |
| 000072590900002 | 000187712000001 | James MA, 1998, HAND CLIN, V14, P1 | Jameson, M. L. 1998. Phylogenetic analysis of the subtribe Rutelina and Revision of the Rutela generic groups (Coleoptera: Scarabaeidae: Rutelinae). Bulletin of the University of Nebraska State Museum 14:1–184. |
| 000072590900002 | 000178148400003 | James MA, 1998, HAND CLIN, V14, P1 | Jameson, M. L. 1998. Phylogenetic analysis of the subtribe Rutelina and revision of the Rutela generic groups (Coleoptera: Scarabaeidae: Rutelinae: Rutelini). Bulletin of the University of Nebraska State Museum 14:1–184. ['‘1997'’]. |
| 000072590900002 | 000171172900018 | James MA, 1998, HAND CLIN, V14, P1 | Jameson, M. L. 1999 (1998). Phylogenetic analysis of the |

| | | | subtribe Rutelina and revisions of the Rutela generic groups (Coleoptera: Scarabaeidae: Rutelini). Bulletin of the University of Nebraska State Museum 14:1–184. |
|---|---|---|---|
| 000072590900002 | 000168258800013 | James MA, 1998, HAND CLIN, V14, P1 | Jameson, M. L. 1998 (1997). Phylogenetic analysis of the subtribe Rutelina and revision of the Rutela generic groups (Coleoptera: Scarabaeidae: Rutelinae: Rutelini). Bulletin of the University of the University of Nebraska State Museum 14:1–184. |
| 000072590900002 | 000086539300009 | James MA, 1998, HAND CLIN, V14, P1 | __________. 1998 (1997). Phylogenetic analysis of the subtribe Rutelina and revision of the Rutela generic groups (Coleoptera: Scarabaeidae: Rutelinae: Rutelini). Bulletin of the University of Nebraska State Museum 14:1–184. |
| 000072590900002 | 000082711000013 | James MA, 1998, HAND CLIN, V14, P1 | James, E.M., Nicholls, A.W., Stemmer, S., Xin, Y., and Browning, N.D., 1998, Proceedings of the 14th International Congress on Electron Microscopy, Vol. 1, p. 209. |
| 000071590000001 | 000241359000097 | Page RN, 1998, J CURRICULUM STUD, V30, P1, DOI 10.1080/002202798183738 | Brin, S., Page, L.: The anatomy of a large-scale hypertextual Web search engine. Computer Networks and ISDN Systems 30(1–7) (1998) 107–117 |
| 000071590000001 | 000240091500045 | Page RN, 1998, J CURRICULUM STUD, V30, P1, DOI 10.1080/002202798183738 | Brin, S., Page, L.: The anatomy of a Large-scale Hypertextual Web search Engine. Computer Networks and ISDN Systems 30(1–7) (1998) 107–117 |
| 000071590000001 | 000185510800036 | Page RN, 1998, J CURRICULUM STUD, V30, P1, DOI 10.1080/002202798183738 | Brin, S., Page, L.: The Anatomy of a Large-Scale Hypertextual (Web) Search Engine. Computer Networks and ISDN Systems, 30(1–7) (1998) 107–117 |
| 000071590000001 | 000171566400006 | Page RN, 1998, J CURRICULUM STUD, V30, P1, DOI 10.1080/002202798183738 | Page, R.D.M. And Holmes, E.C. (1998) Molecular Evolution: a Phylogenetic Approach. Blackwell Science, Oxford, pp.30-31 |
| 000181677600003 | 000263372000008 | Thompson DT, 2003, J TRAVEL MED, V10, P79 | N.Thomas, Imagining minds, Journal of Consciousness Studies 10 (11) (2003) 79–84. |
| 000252591500001 | 000262588100046 | Markell P, 2008, POLIT THEORY, V36, P9, DOI 10.1177/0090591707310084 | Markman M. What is the optimal approach to the administration of intraperitoneal chemotherapy in ovarian cancer? Int J Gyencol Cancer 2008;18(suppl 1):36–9. |
| 000071427200001 | 000243443300005 | Schmidt J, 1998, POLIT THEORY, V26, P4, DOI 10.1177/0090591798026001002 | Schmid CW. 1998. Does SINE evolution preclude Alu function? Nucleic Acids Res 26:4541–4550. |
| 000182704300004 | 000234871900003 | Becker R, 2003, POLIT VIERTELJAHR, V44, P19, DOI 10.1007/s11615-003-0004-5 | Beckstead, A. L. (2003b, Spring). "We're approaching this too narrowly": The need for a broader-based therapy |
| | | | for conflicted, same-sex attracted clients. Division 44 Newsletter, 19, 8–11. |
| 000077273300003 | 000240470700011 | Strahl S, 1998, DEUT MED WOCHENSCHR, V123, P1410, DOI 10.1055/s-2007-1024196 | No references in the letter. |
| 000075466100001 | 000181820100013 | Schannwell CM, 1998, DEUT MED WOCHENSCHR, V123, P957, DOI 10.1055/s-2007-1024104 | Schannwell CM, Schoebel FC, Zimmermann T et al. (2000) Linksventrikuläre diastolische Funktion in der normalen Schwangerschaft. Dtsch Med Wochenschr 123:957–964 |
| 000075466100001 | 000180534800006 | Schannwell CM, 1998, DEUT MED WOCHENSCHR, V123, P957, DOI 10.1055/s-2007-1024104 | Schannwell CM, Schoebel FC, Zimmermann T, Marx R, Plehn G, Leschke M, Strauer BE (2000) Linksventrikuläre diastolische Funktion in der normalen Schwangerschaft. Eine prospektive Untersuchung mittels Mmode Echokardiographie und Doppler-Echokardiographie. Dtsch med Wschr 123:957–964 |
| 000072512500004 | 000181024300001 | Rueger JM, 1998, ORTHOPADE, V27, P72, DOI 10.1007/PL00003481 | No references in the editorial. |

**Table A5. List of double records identified in WoS**

| WoS_UT_Target | WoS_UT_Source 1 | WoS_UT_Source 2 | Comment |
|---|---|---|---|
| 000256997900003 | 000284810400005 | 000284115600005 | |
| 000076975700003 | 000203012100005 | 000239023000005 | |
| 000075085200002 | 000271241600011 | 000270174100011 | The WoS_UT_Source 1 was not covered by the databases of CWTS and iFQ. |
| 000186000600003 | 000221895100009 | 000225772800008 | |